# A Study on the Efficient Product Search Service for the Damaged Image Information

Yonghyun Kim

*School of Computer Science and Engineering, Ulsan National Institute of Science and Technology, Ulsan, Korea, (e-mail: yhkim6087@unist.ac.kr)*

**Abstract:** With the development of Information and Communication Technologies and the dissemination of smartphones, especially now that image search is possible through the internet, e-commerce markets are more activating purchasing services for a wide variety of products. However, it often happens that the image of the desired product is impaired and that the search engine does not recognize it properly. The idea of this study is to help search for products through image restoration using an image pre-processing and image inpainting algorithm for damaged images. It helps users easily purchase the items they want by providing a more accurate image search system. Besides, the system has the advantage of efficiently showing information by category, so that enables efficient sales of registered information.

## 1. INTRODUCTION

Image Retrieval is no longer an unfamiliar concept but has emerged as an important application in computer vision. As it progresses, many researchers have been developing image retrieval, as it can provide users with a high sense of reality and optimal User eXperience (UX) in various fields.

Image Retrieval is especially highly being used in e-commerce. In a research study in ViSenze [1], one of the global visual commerce platforms, according to the survey of 1,515 people in July 2018, around 62% of respondents who are millennials, aged 18-34, mentioned they feel able to search by image is comfortable that enables them to quickly discover the products on their mobile devices. Detailed figures are given in Figure 1. Rather than typing in the information to find the product, taking a picture of the desired product and searching for it directly gives users an intuitive and easy sense of use. This build-up of positive UX soon creates a favorable impression and positive brand value of the search engine (interface). If the search engine, however, does not have high enough accuracy due to low recognition performance for the image, it may have the opposite effect. Therefore, service systems that provide image retrieval should make continuous efforts to achieve and retain good accuracy.

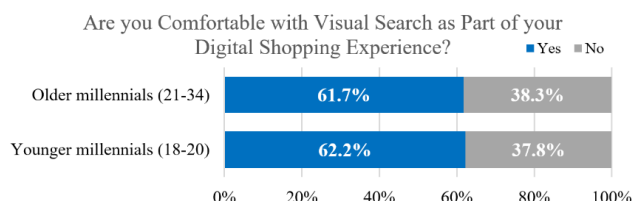

*Fig. 1. Research result of accuracy evaluation (Source: [1])*

The original image retrieval system was implemented by annotating images with metadata, e.g., keywords or captions, manually. But this soon has evolved to be automatically done so that the time and effort spent on annotating be drastically reduced [2]. On the other hand, rather than utilizing metadata, it also has been improving search efficiency by incorporating artificial intelligence technology into CBIR (Content-Based Image Retrieval) concept, which analyzes the images solely by focusing on its content, e.g., colors, shapes, etc. With this steady development of image retrieval, users were able to search for products or objects they found in real life immediately after taking pictures with any camera-mounted portable electronic device.

Although image search technology has evolved, it is still difficult to compare a person's level of awareness with a computer in the accuracy of a search. A research team from Perficient, Inc. [3] has selected 4 search engines that use image recognition, then investigated which search engine performed the best image recognition and compared it with that of humans. The study conducted two different methods: Matching Human Descriptions Evaluation, Accuracy Evaluation. Those results are shown in Figures 2 and 3, separately.

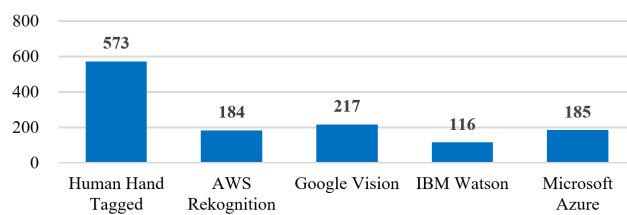

*Fig. 2. Research result of matching human descriptions evaluation: Average score by platforms that how well do generate tags match user descriptions (Source: [3])*

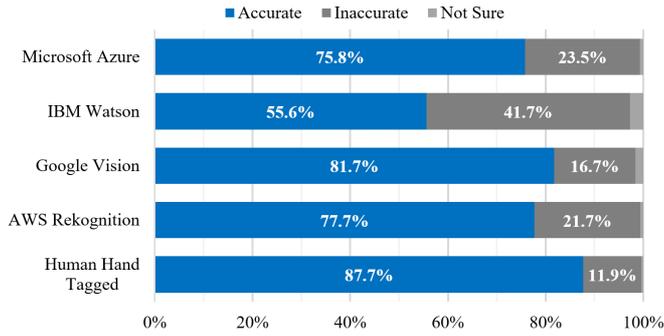

*Fig. 3. Research result of accuracy evaluation: Average matching accuracy (Source: [3])*

Overall, to sum up, Google Vision has the best performance among the four recognition engines, but it still shows that they do not have image recognition through contextual judgments similar to human beings.

This paper presents a new mechanism that improves image recognition accuracy. In general, it is inevitable to capture the photo without having noises, e.g., blocking obstacles or low lightness, on the target even though users try to provide an image with consideration for the machine. If input images are not delivered directly to the machine, and noises are removed as much as possible in advance, the recognition accuracy will increase.

## 2. FUNDAMENTALS AND RELATED WORK

### 2.1 Digital Image Processing

Digital image processing refers to all computer techniques related to digital images. This section introduces some image processing techniques which can use in the pre-processing phase of the proposed product search service.

**Canny Edge Detection.** Edge refers to the part where the pixel value changes rapidly in the image. It can be seen from a mathematical point of view as a rapid increase in the gradient of the pixel value. However, since the 2D image is data in which the discrete values are arranged in a plane, using partial differential to obtain a gradient is impossible. In place of this, the edges of the image can be detected by other techniques that use bitmasks. Among them, Canny Edge Detection [4] is introduced. Canny Edge Detection consists of four computational sequences: Gaussian filtering, gradient calculation, non-maximum suppression, and hysteresis edge tracking.

Firstly, to eliminate noise, apply a Gaussian filter on the image. Gaussian filters are filtering techniques using a bitmask created in an approximate Gaussian distribution function. Since the image to use in our research is 2D data, a filter mask must also be derived from the 2D Gaussian distribution function. The following equation shows a 2D Gaussian distribution function with a mean (0, 0) and standard deviation $\sigma_x, \sigma_y$.

$$G_{\sigma_x \sigma_y}(x, y) = \frac{1}{2\pi\sigma_x\sigma_y} e^{-\left(\frac{x^2}{2\sigma_x^2} + \frac{y^2}{2\sigma_y^2}\right)} \quad (1)$$

Distribution values $G$ derived by substituting the appropriate standard deviation values can express in a 2D array, with the median (middle value) of the array being the highest value due to the characteristics of the Gaussian distribution. This array is called Gaussian Mask and removes noise in the image by using it as a filter mask.

After that, the gradient is calculated by applying a bitmask, i.e., Sobel mask, to the array. Basically, the gradient of $f(x, y)$ is expressed as an equation (2), and its magnitude is expressed as the root mean square of each vector component. However, the multiplication operation is slower than the addition, so it is usually approximated as shown in equation (3) to improve the computational speed.

$$\nabla f(x, y) = \left(\frac{\partial f}{\partial x}, \frac{\partial f}{\partial y}\right) \quad (2)$$

$$\|\nabla f(x, y)\| \approx |f_x| + |f_y| \quad (3)$$

Then, after applying a non-maximum suppression that sets only local maximum pixels to edge from this approximate gradient value, the final edge is selected by going through hysteresis edge tracking by using two thresholds. The sample results of Canny Edge Detection are shown in Figure 7.

**Histogram Stretching.** Histogram of images is the distribution of pixel values in images in bar graph form. It uses the grayscale value for the grayscale image and the color or brightness value for the color image. Among image processing techniques utilizing histograms, histogram stretching increases contrast by extending the histogram to both sides. The formula for histogram stretching for a grayscale image can be expressed as follows.

$$HS(x, y) = \frac{img(x, y) - G_{min}}{G_{max} - G_{min}} \times 255$$

$HS$ stands for histogram stretching, and $img$ represents an input image. $G_{min}$ and $G_{max}$ represent the minimum and maximum value of the grayscale value of the image, respectively. Histogram stretched image samples are shown in Figure 8.

**Histogram Equalization of color image.** In addition to histogram stretching, histogram equalization is one of the algorithms that change the distribution of pixel values in images to be evenly distributed across the entire region. To increase the intensity ratio of color images, there may be a method in which channels are divided into R, G, and B components, and then histogram equalization is performed, respectively, and then merge them. However, because this changes color information, the resulting image can represent a completely different color than the original. Since color information of products can also be used in product search, to increase intensity ratio while preventing changes in color, histogram equalization should be performed only for brightness information. Taguchi [5] proposed a novel method which is combined luminance enhancement and saturation enhancement, in the YCbCr color space. As a result of this application, a sample is presented in Figure 7.

## 2.2 Image Inpainting

Image inpainting is a technology that restores lost images to their original state. In the product search service, image inpainting can be used by considering the obstacle as a mask and inpainting the image when the object to be searched is blocked by the obstacle. Although the image inpainting technique using GANs has emerged recently because of its excellent performance, it is difficult to use GANs in product search services because ground truth images must be existed and be trained in advance. However, NVIDIA's proposed image inpainting technique [6], which makes only the parts of the image to be recovered masked and renormalized, does not require an image of an intact object as an input. Also, due to the processing performance of the irregular mask being superior to the other deep learning-based image inpainting methods, it is appropriate to be used in the product search service system. Figure 4 is an example of applying the technique. If you take a picture of a chair with a cat on it and want to search for information only about the chair, but also you don't want to bother the cat, then this technique can get a cat-disappeared chair picture by just setting the cat as a mask.

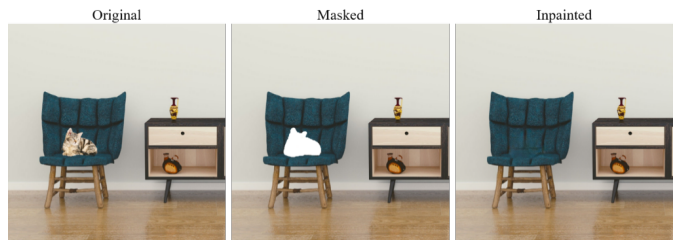

*Fig. 4. Sample result of image pinpointing using partial convolutions*

## 3. EFFICIENT PRODUCT SEARCH SERVICE

### 3.1 Service Structure

The overall service structure [7] consisted of four parts in the manner of the product search service, as given in Figure 5.

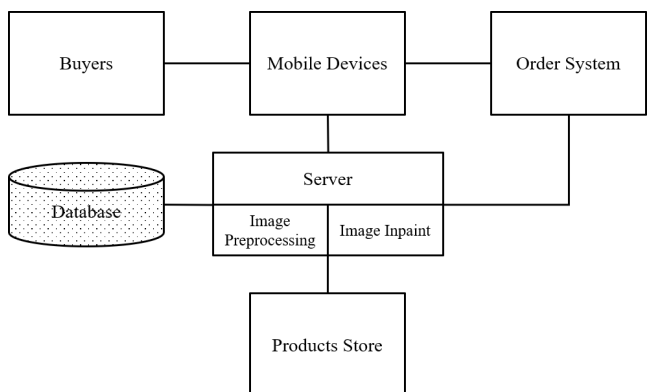

*Fig. 5. Structure of service platform (Source: [7])*

Image retrieval is performed when a buyer uploads images of the desired product to the order system via his or her mobile device. Undamaged images will not be problematic, but damaged or corrupted images will not be properly retrieved. For this, the edges and characteristics of the image are identified through the pre-processing phase. Then, the image is inpainted by going through a deep learning algorithm. After that, metadata is generated from restored images and utilized for serving the most accurate product information which is stored in a database to the buyer. In this paper, the focus was solely on metadata as information for image retrieval, but in reality, information that could better predict the needs of users, i.e., order history or user location, can also be used for searching [7].

### 3.2 Product Search Service

As shown in Figure 6, the structure of the search service system proposed by this study consists of six parts: Image Pre-processing Module, Deep Learning module (image inpainting module), Metadata Generating Module, Image Searching Module, Cluster Analyzing Module (Category Classification Module) and database.

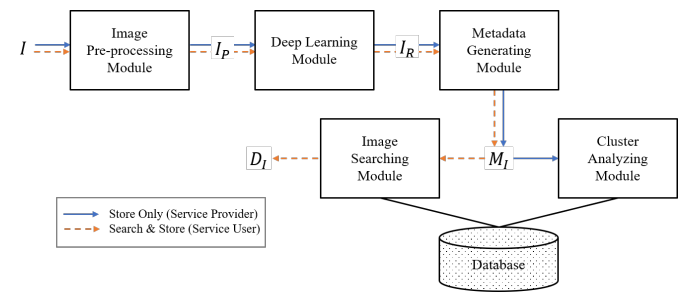

*Fig. 6. Structure of the product search service system*

Image Pre-processing Module modifies input image $I$ to pre-processed image $I_P$ based on colors, edges, and corners through image processing technology which is the same as conventional image search technologies. However, it is important to select appropriate image processing algorithms because choosing the algorithm according to the image type can give a high performance during the next restoration process. More details of this are given in '3.3 Image Pre-processing by Cases.'

Deep Learning Module is a structure that inpaints (restores) the image $I_P$, which is the output data of the previous module. In order to apply to e-commerce, ground truth image (undamaged image) should not be given as input to the search service system. Thus, it will be necessary to apply an inpainting algorithm, i.e., partial convolutions, that uses only a damaged image and mask image for input.

After Deep Learning Module outputs restored image $I_R$, Metadata Generating Module collects metadata of $I_R$ and structures it to store in the database for future comparative analysis. Here, the structured metadata of $I_R$ is $M_I$. Since then, the flow of the system is divided by the status of the service user.

If the service provider is operating the system to store information in the database, $M_I$ goes directly into the Cluster

Analyzing Module. Cluster Analyzing Module clusters, or classifies, similar characteristics by comparing $M_I$ with information stored in the database. As a prerequisite, the database must be categorized from all the image data. After the module determines a category for $M_I$, it then stores in the database so that it belongs to that category.

On the other hand, if the service user who wants to only retrieve images uses the system, $M_I$ generated from the Metadata Generating Module is used as input to the Image Searching Module. Based on $M_I$, this module accesses the database and executes the similarity test, which also includes a category analysis. What is different from the previous case is that the Image Searching Module can obtain product data $D_I$ the most similar to $I$ from the database, and then pass it on to the user. Since then, the Image Searching Module stores $M_I$ in the database to utilize as the potential data.

*3.3 Image Pre-processing by Cases*

This section suggests an image pre-processing mechanism in the product search service system. Good input material for Deep Learning Module can be generated by performing different pre-process in case the input image is color or grayscale.

Firstly, the color image. If the color image provided is blurred and you want to proceed with edge detection, it was found that applying histogram equalization of color first and detecting the edge shows a better result than detecting it from the original image immediately. The sample results are given in Figure 7.

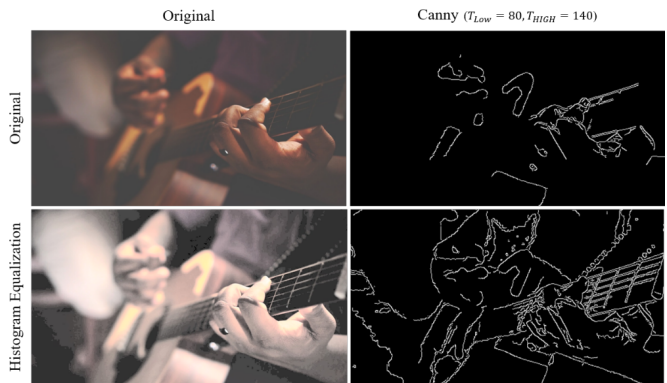

*Fig. 7. Sample results; Left-Top: Original. Left_Bottom: Applied histogram equalization of color from the top. Right: Applied canny edge detection from the left with a low threshold of gradient magnitude ($T_{Low}$)=80 and a high threshold ($T_{HIGH}$)=140*

As you can see in the figure above, histogram equalization of color, which uses image brightness information, has increased the contrast, so that induces the Canny Edge Detector's better performance. An image that went through histogram equalization showed more edges, which has important implications for object recognition, so they produce better results in subsequent image inpainting.

The following is the case for the grayscale image. Grayscale images can be overall improved through sharpening or histogram stretching. As you can see in Figure 8, an unsharpening mask filter focuses on sharpening the edge without controlling contrast. On the contrary, histogram stretching enhanced contrast only. Therefore, it is predicted that the combination of these two will show good performance in image inpainting and product searching if they complement each other's deficiency.

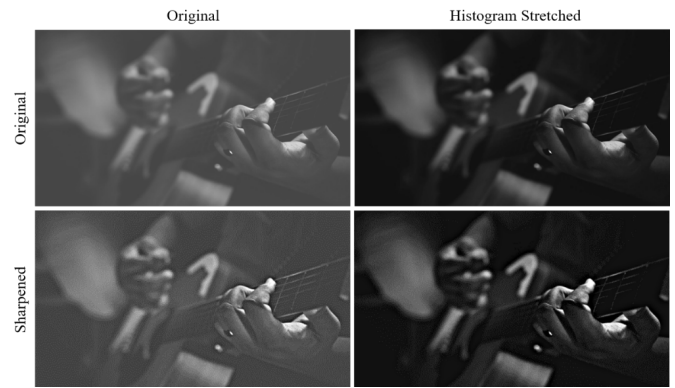

*Fig. 8. Sample results; Left-Top: Original. Bottom: Applied unsharpening mask filter from the top. Right: Applied histogram stretched from the left.*

4. CONCLUSION

In the system proposed in this study, the user can search for the product and its surrounding suppliers with the image of the product they want, thereby making it easier to order the product. Occasionally, if the image or the content of the product is damaged differently from the original, the existing image search system can provide improper information. However, the proposed system helps users to search the desired product well by enabling a highly accurate image search with pre-processing and inpainting. In future work, the author will specifically find out the image preprocessing or inpainting technique optimized for the product search service systems. Then will compare the performance differences between images that have been pre-processed and inpainted with unprocessed by putting them into image search engines. Through this and further research, the author hopes to amplify the accuracy of visual search and further promote image search within e-commerce.